\documentclass[aps,reprint,showpacs,floatfix,shortbibliography,raggedbottom]{revtex4-1}

\draft 
\usepackage[utf8]{inputenc}
\usepackage{amsmath,amsthm,amssymb}
\usepackage{float}
\usepackage{graphicx}
\usepackage{dcolumn}
\usepackage{bm}
\usepackage{color}

\usepackage{hyperref}   
\usepackage[all]{hypcap}
\pdfoutput=1
\DeclareUnicodeCharacter{2212}{-}

\begin{document}
\preprint{APS/123-QED}

\title{Nontrivial low-frequency topological waves at the boundary of a magnetized plasma}
\author{Roopendra Singh Rajawat}
\email{rupn999@gmail.com}
\affiliation{School of Applied and Engineering Physics, Clark Hall, Cornell University}
\author{Tianhong Wang}
\affiliation{School of Applied and Engineering Physics, Clark Hall, Cornell University}
\author{G. Shvets}
\affiliation{School of Applied and Engineering Physics, Clark Hall, Cornell University}
\date{\today}

\begin{abstract}
	The topological properties of a magnetized cold gaseous plasma have recently been explored and the existence of topologically protected edge states has been established. These studies are limited to a magnetized plasma, where ions are infinitely massive and provide a neutralizing background. When ion motion is included, a new class of low-frequency unidirectional topological waves (TSPWs) emerges in the dispersion relation. The group velocity of these waves is in the opposite direction of high-frequency topological electron waves for a given magnetic field direction.  The Berry curvature and Chern numbers are calculated to establish nontrivial topological phase. Additionally, we demonstrate a unique characteristic of ion dominated TSPW propagating above the ion cyclotron frequency: their collisionless damping via coupling to the continuum of lower-hybrid resonant modes localized inside a smooth plasma-vacuum interface. These finding broadens the possible applications and observations of these exotic excitations in space and laboratory plasmas.
\end{abstract}
\maketitle

{\it Introduction.-} 
Independently of their state of matter -- solid~\cite{Hasan_rmp_2010,Haldane_prl_2008,soljacic_prl_2008}, gaseous and fluid~\cite{delplace_science_2017,Yang_prl_2015}, or plasma~\cite{Parker_prl_2020b,fu_nc_2021} -- a wide range of materials is found to exhibit non-trivial topological properties that fundamentally impact wave propagation at domain walls between topologically-distinct bulk materials. Specifically, bulk materials possessing propagation bandgaps (i.e., acting as ``insulators" for wave-like polaritonic perturbations of certain energies) and lacking the time-reversal (TR) symmetry can be characterized by an integer invariant, known as the Chern number~\cite{Ozawa_rmp_2019}, assigned to every bandgap. Bulk-edge correspondence (BEC) -- originally established in condensed matter physics~\cite{Klitzing_prl_1980,Hatsugai_prl93,Hasan_rmp_2010} and later extended to photonics and metamaterials, cold atomic gases, and classical fluids \cite{Haldane_prl_2008,soljacic_prl_2008,Silveirinha_prb_2015,Gangaraj_prl_2020,Goldman_np_2016,tauber_jfm_2019}, -- predicts the existence and number of gapless unidirectional edge states that are spectrally-located within a common bandgap of the two bulk materials sharing a domain wall.


Recently,  nontrivial topology of high-frequency edge states (surface plasmon polariton; SPP's), occurring at the boundary of a cold magnetized plasma, have been established \cite{Gao_nc_2016,yang_nc_2016,Gangaraj_prb_2019,Pakniyat_ieee_2020,Parker_prl_2020a, Parker_prl_2020b,Parker_jpp_2021,fu_nc_2021,fu_jpp_2022,Qin_arxiv_2022} for gyrotropic plasmonic materials \cite{Gao_nc_2016, yang_nc_2016, Gangaraj_prb_2019,Pakniyat_ieee_2020}, and continuum plasma fluids \cite{Parker_prl_2020a, Parker_prl_2020b,Parker_jpp_2021,fu_nc_2021,fu_jpp_2022,Rajawat_arxiv_2022a,Qin_arxiv_2022}.   It has been shown that for certain choices of plasma density ($n$),  stationary external magnetic field (${\bf B_0} =  B_0 \hat{z} $) and parallel wavenumber $k_z$,  SPP's can exist at the plasma-vacuum \citep{Parker_prl_2020b,Parker_jpp_2021} or plasma-plasma interface\cite{fu_nc_2021,fu_jpp_2022}, provided both mediums have different topological phase .  

Study of nontrivial topology of low-frequency edge states is still in infancy. In plasmonic materials, the low-frequency SPPs ceases to exist due to typical Drude behavior and resulting dissipation. One can however make an artificial material out of very thin metallic wires, which supports low-frequency SPPs \cite{Pendry_prl_1996,Shelby_science_2001,Gangaraj_prr_2020} and removes dissipation. In continuum fluids, like a plasma, low-frequency electrostatic \cite{Emeleus_pl_1966,Sook_pop_1999} and electromagnetic  \cite{Cramer_ps_1995, Boris_prl_2000} surface waves arises naturally due to ion motion and exist well below plasma frequencies. For a cold magnetized plasma,  low-frequency electromagnetic waves \cite{Cramer_ps_1995, Boris_prl_2000, Parker_prl_2020a} have been studied, however, nontrivial topology of low-frequency waves has not been discussed. In a recent work, the nontrivial topology of a magnetized fusion plasma in the Alfven continuum has been discussed in the presence of magnetic shear\cite{Parker_prl_2020a} and found unidirectional reversed-shear Alfven eigenmodes. These low-frequency modes are non-dissipative and exist at the neutral layer of magnetic shear.

In this letter, without going into much complexity, we use a simplified cold magnetized plasma slab model to observe the effect of ion motion on nontrivial topological edge states, occurring at the plasma-vacuum interface, and calculate Chern numbers of energy bands for mobile ions. We discuss the limitation on parallel wavenumber $k_z$ for a chosen value of ion mass, numerically obtain the dispersion relation of low-frequency topological edge sates, demonstrate their existence and topological protection by exciting an edge state enroute to a rectangular discontinuity using a full 3D particle-in-cell simulation code.

We consider the cold-plasma slab model of a magnetized, stationary plasma, where electron and ion collisions are neglected. The external magnetic field is uniform (${\bf B_0} = B_0 \hat{z}$) and is kept in $\hat{z}$ direction. The linearized set of governing equations for an infinite homogeneous plasma is
\begin{eqnarray} \label{eqset}
\frac{\partial {\tilde{\bf v}_e}}{\partial t} & = & -{\omega_{pe} {\bf E}} - \Omega_{e} { \tilde{\bf v}_e} \times \hat{z},  \\
\frac{\partial {\tilde{\bf v}_i}}{\partial t} & = & { \omega_{pi} {\bf E}} + \Omega_{i} {\tilde{\bf v}_i} \times \hat{z},   \\
\frac{\partial {\bf E}}{\partial t} & = &  \nabla \times {\bf B} + \omega_{pe}  \tilde{\bf v}_e - \omega_{pi} \tilde{\bf v}_i,  \\ 
\frac{\partial {\bf B}}{\partial t} & = & - \nabla \times {\bf E}.
\end{eqnarray}
where ${\bf v}_e$ and ${\bf v}_i$ are the electron and ion fluid velocities, $\bf E$ the electric field, ${\bf B}_0$ the external stationary background magnetic field, $\bf B$ the perturbation magnetic field, $|e|$ the charge on an electron, $m$ and $M$ are the mass of electron and ion, $\alpha = m/M$, $c$ the speed of light, $\omega_{pe} = \sqrt{4 \pi n_e e^2/m}$ the electron plasma frequency, $\omega_{pi} = \sqrt{4 \pi n_i e^2/M}$ the ion plasma frequency, $\Omega_{i} = e B_0/M$ the  ion cyclotron frequency and $\Omega_{e} =  e B_0/m$ the electron cyclotron frequency. We have used renormalized velocities $\tilde{\bf v}_e = {\bf v}_e/\omega_{pe}$ and $\tilde{\bf v}_i = {\bf v}_i/\omega_{pi}$, reference electric field $E_0$, and reference frequency $\omega_{p0}$. The time is normalized to ${\omega_{p0}}^{-1}$,  frequency to $\omega_{p0}$, length to $k_p^{-1} = c/\omega_{p0}$, electric field to $E_0$ and magnetic field to $E_0/c$, electron and ion velocities to $eE_0/m$ and $eE_0/M$, respectively.

Using Fourier transformation $\partial_t \rightarrow - i \omega$ and $\nabla \rightarrow i { (k_x, k_y, k_z)}$, we obtain the eigenvalue equation $H \mid \psi \rangle = \omega \mid \psi \rangle$,  where $\mid \psi \rangle = \left( {\bf v_e \, v_i \, E \, B} \right)^{T}$ is a 12-element eigenvector , and $H$ is a $12 \times 12$ Hermitian matrix and given by 
$$ 
H = \begin{pmatrix}
-i \Omega_e \hat{z} \times & 0 & -i \omega_{pe} & 0 \\
0 & -i \Omega_i \hat{z} \times & i \omega_{pi} & 0 \\
i \omega_{pe} & -i \omega_{pi} & 0 & -{\bf k \times} \\
0 & 0 & {\bf k \times} & 0
\end{pmatrix}
.
$$
The dispersion relation is obtained by solving equation $det \left( H - \omega {\bf I}_{12} \right) = 0$. For each nonzero value of $k_z$, the system has 12 eigenvalues $\omega_{\pm n}$, where $n = 1, 2, 3, 4, 5$ is the index of bands. Note that, we have two zero frequency bands $\omega_{0} = 0$. If thermal corrections are introduced, zero frequency bands become positive/negative frequency slow magnetosonic modes in the limit $k_\perp = 0$ \cite{Stix}. We change the problem into 2D by fixing $k_z$ and consider a parameter space $k_\perp = (k_x^2 + k_y^2)^{1/2}$. Since the problem is isotropic in the plane perpendicular to the magnetic field, we keep $k_x = 0$. All bandgaps closes at $k_z = 0$, however, a gap always exist between band 1 and 2 for a non-zero value of $k_z \neq 0$. At a fixed $k_z$, in the limit $k_\perp  \rightarrow \infty$,  $(\omega_1, \omega_2, \omega_3, \omega_4, \omega_5) \rightarrow (0, \omega_{LH}, \omega_{UH},ck_\perp,ck_\perp)$, where $\omega_{UH} = \left( \omega^2_{pe} + \Omega^2_e \right)^{1/2}$ is the upper-hybrid resonance frequency and  $\omega_{LH} =  \left(  \Omega_i^2 +  \frac{\Omega_e^2 \omega_{pi}^2 }{\omega_{UH}^2} \right)^{1/2} $ is the lower-hybrid resonance frequency.

\begin{figure} \label{Fig1}
    \centering
 \includegraphics[width=1.\linewidth]{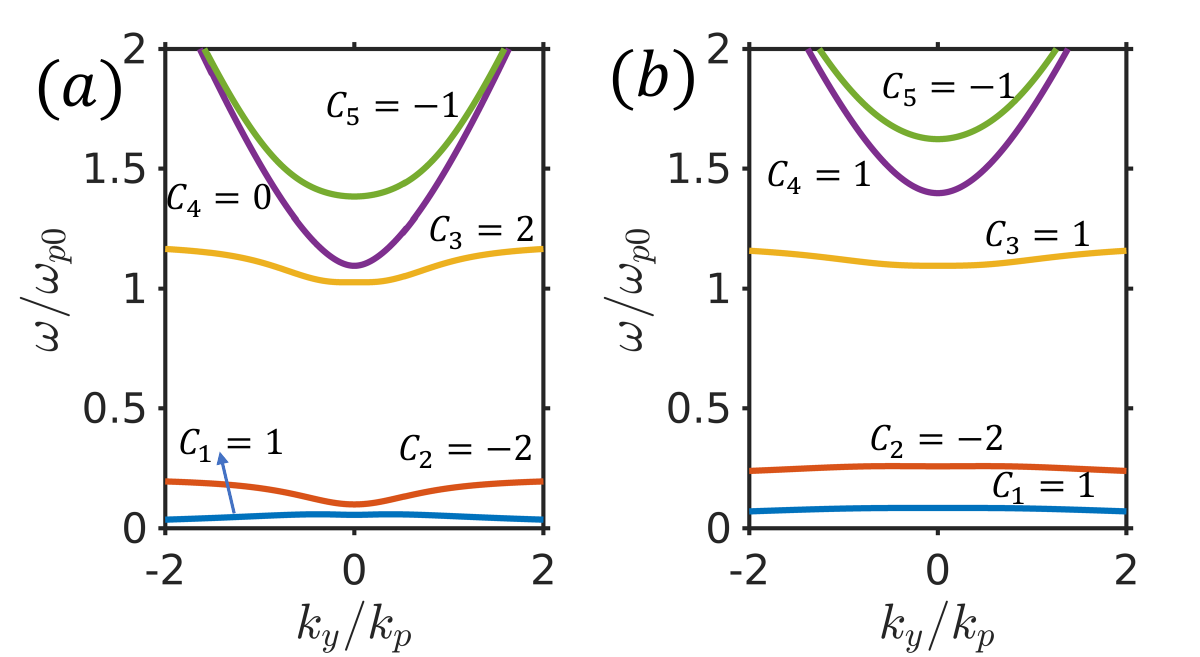}
        \caption{Spectrum of a magnetized, homogeneous, cold plasma as a function of $k_y$, when both electron and ion species are mobile. The spectrum is plotted for parameters $\Omega_{e}/\omega_{p0} = 0.5, \, M/m = 5$, (a) $k_z/k_p = 0.4 $, (b) $k_z/k_p = 1.0$. The Chern numbers for positive frequency bands are shown. 
        }            
\end{figure}  

The topology of a band gap is characterized by its gap Chern number \cite{Hatsugai_prl_1993} , which is defined by $ C_{i,i+1} = \Sigma_{n=-5}^{i} C_n$. The Berry curvature and Chern number are calculated as described in \citep{Parker_prl_2020b, Parker_jpp_2021, fu_nc_2021}. The Chern numbers associated with bands 1, 4, and 5 are integers, however, Chern numbers for bands 2 and 3 are not integers, which is a result of an insufficient smooth linear operator at small scales in continuum fluids that does not compactify the space in finite Brillouin zone \cite{Silveirinha_prb_2015}. To compactify Brillouin zone in a fluid continuum and to ensure that Chern numbers are integers, we use same regularization strategy as discussed in \cite{Silveirinha_prb_2015,Parker_prl_2020b, Parker_jpp_2021}. We choose $\omega_{p,s} = \omega_{p,s}/ \left( 1 + k_\perp^2/k_c^2 \right)$, where $k_c$ is a very big cutoff wavenumber, and subscript $s=e,i$ stands for species electron and ion. The positive frequency bands spectrum $\left(\omega,k_y \right)$ for parameters $\Omega_e/\omega_{p0} = 0.5, \, \Omega_i/ \omega_{p0} = 0.1, \, M/m = 5$ and two different value of parallel wavenumbers $k_z/k_p = 0.4, \, 1$ are shown in Fig. \ref{Fig1}(a,b), respectively. We have also shown band Chern numbers obtained using $k_c = 100$. The Chern number for each positive frequency band is $\left( C_{1}, C_{2}, C_{3}, C_{4}, C_{5} \right)$ = $(1, -2, 2, 0,1)$ for parameters $k_z/k_p =0.4,  \Omega_e/\omega_{p0}=0.5,  M/m=5$ and $\left( C_{1}, C_{2}, C_{3}, C_{4}, C_{5} \right)$ = $(1, -2, 1, 1,-1)$ for parameters $k_z/k_p =1.0,  \Omega_e/\omega_{p0}=0.5,  M/m=5 $. The gap Chern number for a low frequency band gap is $C_{1,2} = 1$ and for a high frequency band gap is $C_{2,3} = -1$. Thus multiple bands are topologically nontrivial and can support multiple topological gapless edge states. Note that, as discussed in \cite{Parker_prl_2020b}, bands 3 and 4 touch at $k_z^{*} = \Omega_e /\sqrt{1+\Omega_e}$. Since band gap closes at $k_z^{*}$, we observe different Chern numbers for bands 3 and 4 for wavenumber ranges $k_z < k_z^{*}$ and $k_z > k_z^{*}$ as shown in Fig. \ref{Fig1}(a) and \ref{Fig1}(b), respectively. Interestingly, the Chern number of low-frequency bands (band 1 and 2) are universal and do not change with changing the parallel wavenumber or finite ion mass. Notwithstanding, changing the magnetic field direction flips the sign of the Chern numbers. We have calculated Chern numbers for various ion mass ranges $M/m = 5-18360$, and found that ion mass does not affect the Chern number of any band.  Surprisingly, the Chern number for band 1 ($C_{1} = 1$) agrees with the Chern number obtained by earlier study \cite{Parker_prl_2020a}. For infinitely massive ions, the Chern numbers for bands $\omega_{\pm 1}$ become zero (trivial) and for bands $\omega_{\pm2}$ the Chern number becomes $\mp 1$, respectively, which is exactly same result earlier studies obtained for single mobile specie plasma \cite{Parker_prl_2020b,Parker_jpp_2021,fu_nc_2021,fu_jpp_2022}.  

\begin{figure} \label{Fig2}
    \centering
 \includegraphics[width=1.\linewidth]{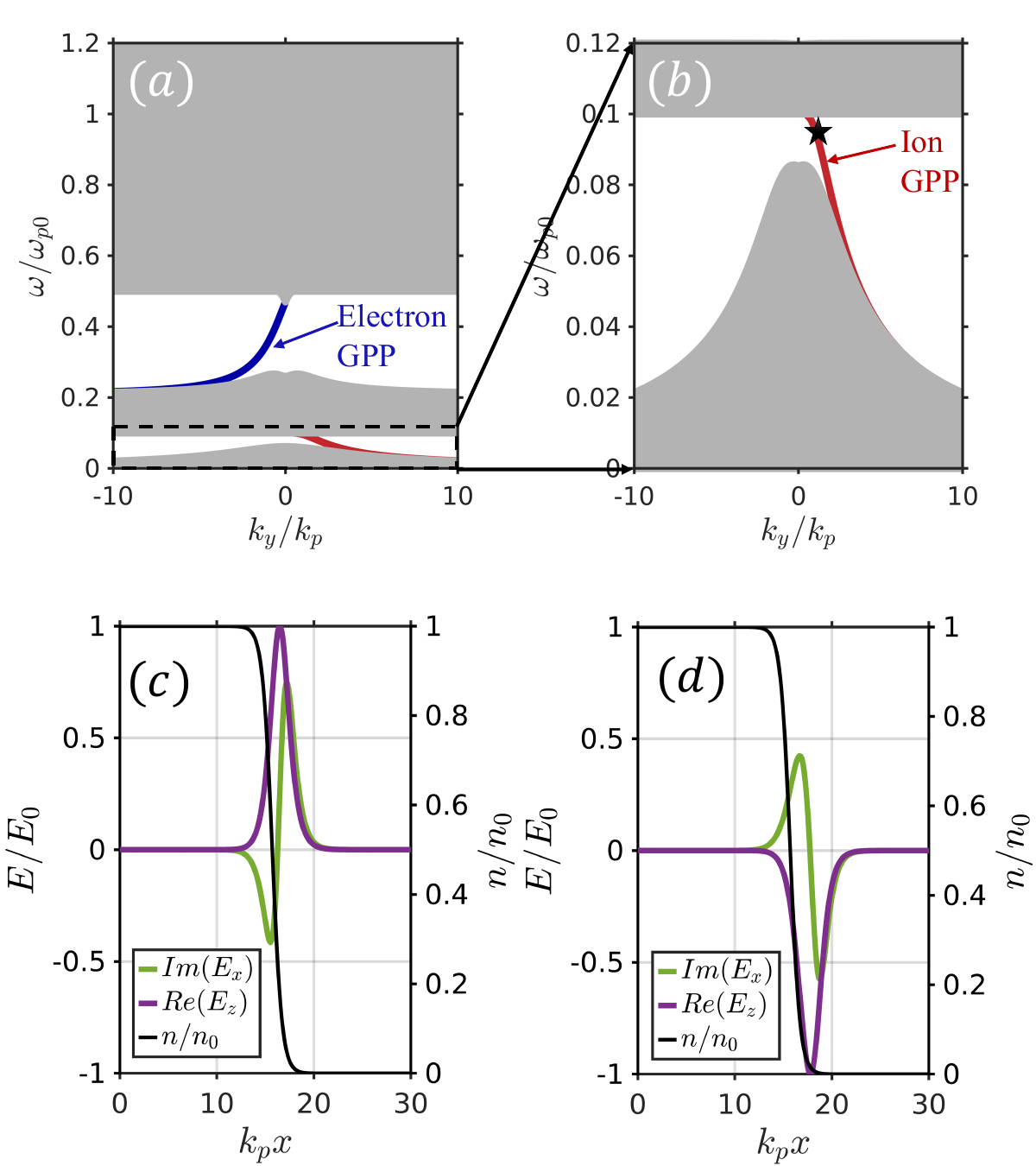}
        \caption{ (a) The band spectrum $\omega(k_y)$ with two band gaps is shown for parameters $\Omega_{e}/\omega_{p0} = 0.5, \, k_z/k_p = 1.0, \, M/m = 5, \, k_p \delta l = 1$. The gray color lines show bulk modes. The high (electron dynamics dominated) and low frequency (ion dynamics dominated) topological edges states are shown by blue and red color lines, respectively.(b) Zoom in view of low-frequency modes. (c) The non-zero eigenvectors (left $y-$scale) for a high frequency mode at $k_y/k_p = -1$. (d) The non-zero eigenvectors (left $y-$scale) for a low frequency mode at $k_y/k_p = 1$. The density profile of plasma in $x-$direction is shown in (c,d) by a black color line (right $y-$scale). 
        }            
\end{figure}  

Bulk-edge correspondence states that an edge mode exists in the common band gaps at the interface between two topologically distinct materials with different gap Chern numbers. We have identified all existing topological nontrivial band gaps in a band spectrum. To prove existence of gapless edge states between band gaps, we use a numerical eigenvalue solver \cite{fu_nc_2021} to solve a set 1D differential equations (1) - (4) in an inhomogeneous plasma by replacing $x-$coordinate dependent electron and ion plasma frequencies $\omega_{pe} \rightarrow \omega_{pe}(x)$ and $\omega_{pi} \rightarrow \omega_{pi}(x)$, respectively. The background magnetic field is uniform (${\bf B} = B_0 \hat{z} $) and constant in z-direction. The plasma density is nonuniform only in the $x-$direction, shown in Fig. \ref{Fig2}(c,d) by a black color line. We choose the density profile $n(x) = \frac{n_0}{2} \left( 1 + tanh \left[ \left( x_0 - x\right)/\delta l  \right] \right) $, where $n_0$ is a maximum density of plasma, $x_0$ the location of the interface and $\delta l$ is width of the interface. We use artificial conducting boundary conditions at $x = 0$ and $x = l$.  The artificial conducting boundary condition yields a nonphysical mode which is uninteresting and not shown here. 

\begin{figure} \label{Fig3}
    \centering
 \includegraphics[width=0.5\linewidth]{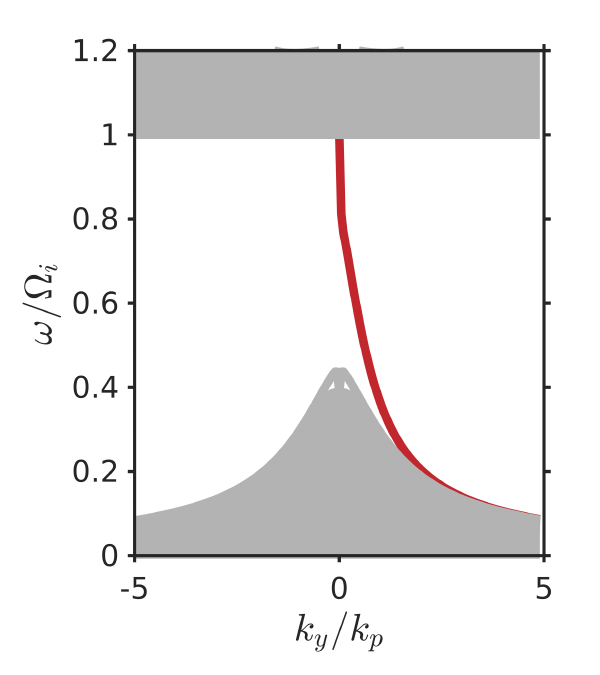}
        \caption{ The band spectrum $\omega(k_y)$ for low frequency modes for the parameters  $\Omega_{e}/\omega_{p0} = 0.5, \, k_z/k_p = 0.01; \, M/m = 1836$.
        }            
\end{figure}

It has been already established \citep{Parker_prl_2020b, Parker_jpp_2021, fu_nc_2021} that for a magnetized plasma an edge state arises between bands 2 and 3, and can also be confirmed in our results as shown by the blue color line in Fig. \ref{Fig2}(a). This high-frequency edge state is dominated by electron dynamics, hence we name it electron topological surface plasma waves (ETSPW). ETSPWs are localized near the center of the density profile which is confirmed by Fig. \ref{Fig2}(c), where the centroid of non-zero components of electric field (eigenmodes), obtained using a numerical eigenvalue solver, is localized near the center of the density profile. Nonetheless, we have another band gap exist between band 1 and 2 and the gap Chern number for this gap is $C_{1,2} = -1$. According to the bulk-boundary correspondence, a gapless edge state must exist in this band gap. We present that a low-frequency edge state exists between bands 1 and 2 for a finite ion mass as shown by the red color line in Fig. \ref{Fig2}(b). This edge state is mostly governed by ion dynamics, hence we name it ion topological surface plasma wave (ITSPW). The ITSPWs are localized near the plasma-vacuum interface, which is confirmed by Fig. \ref{Fig2}(d), where the centroid of non-zero components of electric fields is localized near the interface. Interestingly, the sign of group velocity of low-frequency edge states is in the opposite direction of high-frequency waves, {i.e.}, for a given magnetic field direction, ETSPWs and ITSPWs will propagate in opposite directions. The ITSPWs unidirectional frequency window is confined between band 1 and ion cyclotron resonance frequency. The LH resonance frequency bulk bands exist from ion cyclotron resonance frequency to band 2.  The Band 1 at $k_\perp= 0$ limit becomes left-hand-circularly polarized (LCP) mode and its dispersion relation is given by \cite{Stix}.
\begin{equation} \label{LCP} 
k_z^{LCP}  =  \omega \sqrt{ 1+ \alpha+ \frac{1}{( \Omega_i - \omega)( \Omega_e + \omega)}}.
\end{equation}
The second band gap remains open for every finite value of $k_z$ and closes only at $k_z = \infty$, however, in the limit $k_z \rightarrow \infty$, band gap becomes infinitesimally small. A finite band gap can be tuned using equation \eqref{LCP} at an appropriate frequency value $\omega < \Omega_i$. It should be noted that higher the mass of an ion, lower would be the value of $k_z$ to open a finite band gap to obtain unidirectional frequency window. One needs to use a very small value of $k_z$ to open a finite band gap for higher massive ions. For example, to open a unidirectional frequency window of $\Delta \omega = 0.2 \Omega_i$ for mass ratios $M/m = 5$ and $M/m = 1836$, one needs $k_z/k_p = 0.75$ and $k_z/k_p = 0.042$, respectively. The dispersion for a realistic mass ratio $M/m = 1836$ is shown in Fig. \ref{Fig3}(a) which is plotted for parameters $k_z/k_p = 0.01, \Omega_e/\omega_{p0} = 0.5, M/m = 1836$ where red color line shows a gapless edge mode and gray color lines represent bulk modes. It is worth mentioning that for higher massive ions, edge states become plane waves. The higher ion mass will also yield slower frequency waves. 

\begin{figure}\label{Fig4}
    \centering
 \includegraphics[width=1.\linewidth]{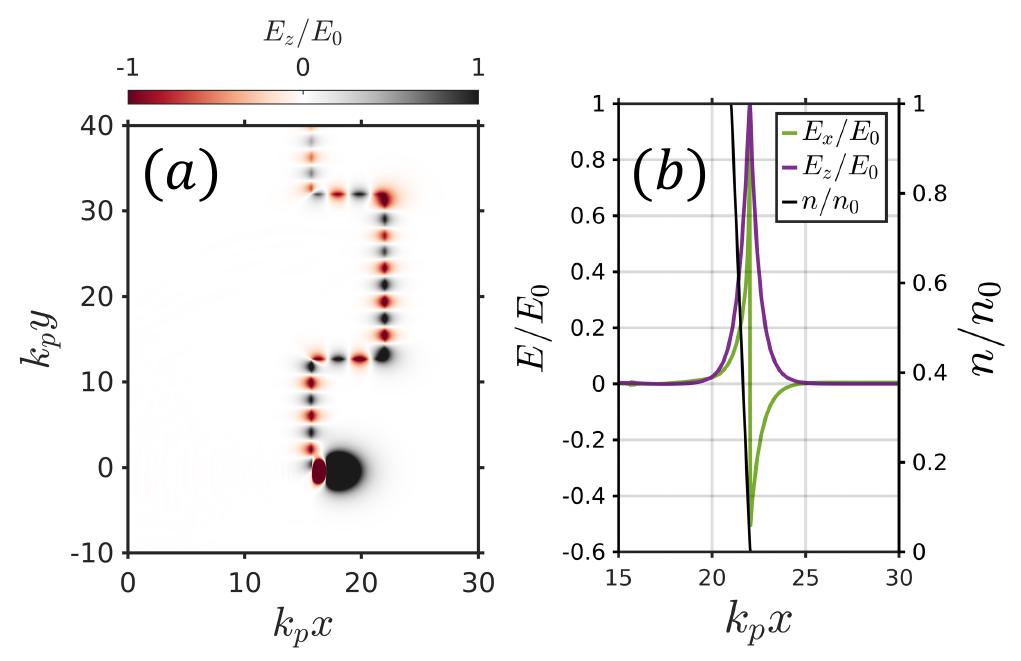}
        \caption{The simulation is carried out for parameters $\, k_z/k_p = 1.0, \Omega_{ce}/\omega_{p0} = 0.5, \, M/m = 5, \, k_p \delta l = 1.0$. (a)  A time snapshot of normalized intensity of electric field $E_z$ at $k_p z = \pi$. (b) The $x-$direction variation of normalized electric fields (left scale) $E_x$ (green line) and $E_z$ (violet color) at $k_p y = 4 \pi$. The density profile as a function $x-$coordinate (black line; left scale). To save computational time, we have considered incompressible plasma fluid, which does not give any nonlinear effects.
        }            
\end{figure}

{\bf 3D particle-in-cell simulation:-} We demonstrate existence and verify their topological protection using a 3D particle-in-cell simulation code\cite{pukhov_1999}. The plasma density profile is shown in Fig. \ref{Fig4}(c), which varies linearly (ramp) from  $n_{0}$ to $0$ within $ k_{p} \delta l =1.0$ length. A $z-$direction periodic dipole source chain is kept at $(x, \, y) = (17, \, 0)k^{-1}_{p}$, which excites propagating edge modes in $-yz \, (+yz)$ plane for positive (negative) external magnetic field ($\bf B_0$). The simulation box size is $L_{x} \times L_{y} \times L_{z} = 30 k^{-1}_{p} \times 50 k^{-1}_{p} \times 6.28 k^{-1}_{p}$. The spatial and temporal resolutions are $\Delta x = 0.02 k^{-1}_{p}$, $\Delta y = 0.2k^{-1}_{p}$, $\Delta z = 0.25 k^{-1}_{p}$ and $\Delta t = 0.8 \Delta x$, respectively. We have used 4 particles per cell. The simulation box has periodic boundaries in the $z-$direction for both particles and electromagnetic fields, absorbing in $x-$ and $y-$directions for electromagnetic fields and reflecting for particles. A rectangular discontinuity is kept at $k_p y = 12$. 

The ITSPW mode is excited for the wavenumber $k_y/k_p = 1$ as shown in Fig. \ref{Fig2}(b) by a black star. The mode frequency and wavenumber $k_y$ obtained from the simulation results agrees with the theoretical dispersion relation. It is clear from Fig. \ref{Fig4}(a,b) that edge states are unidirectional, robust and propagate without any defect and scattering around the discontinity. The centroid of the edge state is localized near the vacuum as shown in Fig. \ref{Fig4}(c), which agrees with the theory. 

{\bf Damped ion topological surface plasma waves:-} Now, we illustrate a distinctive feature of ion-dominated TSPWs propagating above the ion cyclotron frequency: their collisionless damping occurs through coupling with the continuum of lower-hybrid resonant modes, which are localized within a smooth plasma-vacuum interface.. We construct the dispersion relation for damped and undamped topological surface plasma waves (TSPW's) localized near the plasma-vacuum interface. The applied magnetic field ($ {\bf B_0} = B_0 \hat{z}$) is uniform in space and constant in time. We assume that electromagnetic fields are harmonic, i.e., $\propto e^{i k_y y + i k_z z - i \omega t}$, where $k_y$ and $k_z$ are the wavenumbers in the $y$- and $z$-directions and $\omega$ is the wave frequency. After the Fourier transformation, $\partial_t \rightarrow - i \omega; {\bf \nabla} \rightarrow \left( \partial_x, i k_y, i k_z \right)$, Maxwell equations \cite{Jackson} can be written as
\begin{align}
{\bf \nabla} \times {\bf E} & =  i k_0 {\bf B}, \label{eq1}\\
 {\bf \nabla } \times {\bf B} & =  -i k_0 \hat{\epsilon} {\bf E}, \label{eq2}
\end{align}

where the dielectric tensor $\hat{\epsilon}$ for a cold magnetized plasma is given by \cite{Stix}
\begin{equation*}
  \epsilon = 
\begin{pmatrix}
\epsilon_t & - i \epsilon_{g} & 0  \\
i \epsilon_{g} & \epsilon_{t} & 0 \\
0 & 0 & \epsilon_{a}
\end{pmatrix}.
\end{equation*}

Here $ \epsilon_{t}  =  1 - \frac{\omega^{2}_{pe} }{ \omega^{2} - \Omega^{2}_{e}} - \frac{\omega^{2}_{pi} }{ \omega^{2} - \Omega^{2}_{i}} ,  
    \epsilon_{g}  = - \frac{\Omega_{ce}}{\omega}  \frac{\omega^{2}_{pe}}{\omega^{2} - \Omega^{2}_{e}} + \frac{\Omega_{i}}{\omega}  \frac{\omega^{2}_{pi}}{\omega^{2} - \Omega^{2}_{i}},  
    \epsilon_{a}  =  1 - \frac{\omega^{2}_{pe}}{\omega^2 } - \frac{\omega^{2}_{pi}}{\omega^2 } $, $ \bf E$ is the Electric field, $\bf B$ the magnetic field, $k_0 = \omega/c$, $c$ the speed of light, $\omega_{c} = e {B_0} /m$ the electron cyclotron frequency, $\omega_{p0}  = \sqrt{4 \pi n_{0} e^{2}/m}$ the plasma frequency for the bulk density $n_{0}$, $-e$ and $m$ are the electric charge and mass of an electron. After linearization, a set of coupled first order differential equations having the electromagnetic field components $B_x$, $B_y$, $B_z$, and $E_z$ is written as

\begin{equation} \label{eq3}
  \frac{\partial \mathbf{\psi}}{\partial x}   = -\frac{i}{\epsilon_t(x)} M \mathbf{\psi}, \ \ \ \mathbf{\psi}  = \left(B_x,B_y,B_z,E_z\right)^T
\end{equation}
where 
\begin{equation} \label{eq4}
M = \begin{pmatrix}
0 & k_y \epsilon_t & k_z \epsilon_t & 0 \\
-k_y \epsilon_t & 0 & 0 & k_0 \epsilon_a \epsilon_t \\
\frac{k_0^2 \epsilon_{tg}^2 - \epsilon_t k_z^2}{k_z} & -ik_z \epsilon_g & ik_y \epsilon_g & - \frac{k_0 k_y}{k_z}  \epsilon^2_{tg} \\
ik_0 \epsilon_g & \frac{\epsilon_t^2 k_0^2 - k_z^2}{k_0} &  \frac{k_0 k_y}{k_z}  & -ik_y \epsilon_g
\end{pmatrix},
\end{equation}
where $\epsilon_{tg}^2 \equiv \epsilon^2_t - \epsilon_g^2$.

The field components $E_x$ and $E_y$ has algebraic relationship as
\begin{align}
E_x & = \frac{1}{\epsilon_t} \left[ - \frac{k_y}{k_0 k_z} \left( -i k_0 \epsilon_g E_z + k_z B_z \right) + \frac{k_z}{k_0} B_y - i \frac{k_0}{k_z} \epsilon_g B_x  \right], \label{eq7} \\
E_y & = \left( \frac{k_y}{k_z} \right) E_z - \left( \frac{k_0}{k_z} \right) B_x.  \label{eq8}
\end{align}

For simplicity, we consider a monotonically decreasing plasma density ramp, i.e., $n(x) = (1 - x/\delta l)$. To make further analytic progress in understanding this scaling with $\delta l$, as well as the properties of the ITPSW quasi-modes, we assume that the electromagnetic fields $\mathbf{\psi}$ are nearly-uniform inside the SPI. Then, their increment $\Delta \mathbf{\psi} \equiv \mathbf{\psi}^{v} - \mathbf{\psi}^{p}$ (where $p$ ($v$) superscripts label the fields in the plasma (vacuum) regions) can be estimated as $\Delta \mathbf{\psi} \approx i M \mathbf{\psi} \left( \int^{\delta l}_{0} dx/\epsilon_t(x) \right)$. After expanding $\epsilon_t(x)$ in the vicinity of $x=x_{\rm LH}$ as $\epsilon_{t} \approx \tilde{x}/ \tilde{\delta l}$, where $\tilde{\delta l} = \delta l - x_{\rm LH}$ is the distance of the local lower-hybrid resonance point from the edge of the plasma and $\tilde{x} = x - x_{\rm LH}$, the integral over the pole is calculated to be approximately equal to $- i \pi \tilde{\delta l}$. To leading order in $\delta l$, we find that $B_x$ and $B_y$ are continuous ($B^p_{x,y} - B^v_{x,y} \approx 0$) while $B_z$ and $E_z$ change across the interface: $ k_z \left( B^p_{z} - B^v_{z} \right) \approx  -i \pi \tilde{\delta l} \epsilon_g^* R$ and $ k_0 \left( E^p_{z} - E^v_{z} \right)  \approx  - \pi \tilde{\delta l}R$. Here $R = k_y \left( k_z B_z - i \epsilon_g^*(x) E_z \right) - k_z^2 B_y + ik_0^2 \omega_g^*(x) B_x$ is calculated on the vacuum side of the interface and $\epsilon_g^* = \frac{- \omega \Omega_e}{\Omega^2 _{e} \omega^2_{pi} - \omega^2_{pe} \left( \omega^2 - \Omega^{2}_{i} \right) }$.

Using these boundary conditions, a dispersion relation $D_{\ast}(\omega,\tilde{k}_y) = 0$ for the edge states can be expressed as follows:
\begin{eqnarray}
& \left(\kappa_1 + \kappa_0 + Q_1 \right) \left( P_2 - k_y \right) =  \left(\kappa_2 + \kappa_0 +   Q_2 \right) \left( P_1 - k_y \right)\label{eq:dispersion}
\end{eqnarray}

where $\kappa_0$ and $\kappa_{1,2}$ are the inverse decay lengths in vacuum and plasma, respectively, and remaining coefficients are given as $Q_1 = - \pi \delta_l k_y \frac{\omega_c}{\omega} (\alpha_1 + s_1 \alpha_2)$, $Q_2 = - \pi \delta_l k_y \frac{\omega_c}{\omega} (\alpha_1 + s_2 \alpha_2)$, $P_1 = p_1 + i s_1 \kappa_0 + \frac{\omega}{\omega_c}\frac{\kappa_1 + \kappa_0}{k_y}$,  $P_2 = p_2 + i s_2 \kappa_0 + \frac{\omega}{\omega_c}\frac{\kappa_2 + \kappa_0}{k_y}$, $\alpha_1 = i \left( k_0 \epsilon_g^* - k_y (-k_y \epsilon_g^*(x) + \kappa_0) \right)$ and $\alpha_2 =  -(k^2_y + k^2_z + k_y \kappa_0 \epsilon_g^*(x))$. See supplemental for details. 

\begin{figure} \label{Fig5}
\includegraphics[scale=0.6]{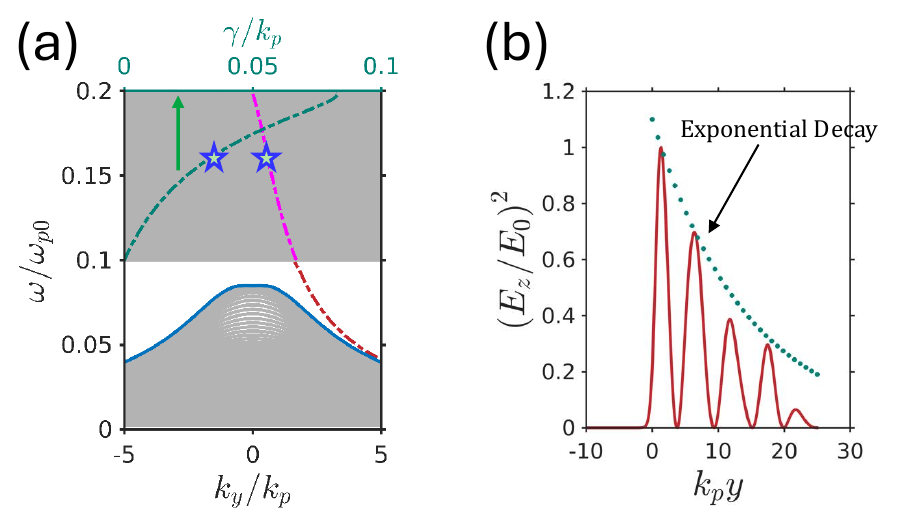}
\caption{(a) Propagation band spectrum for a semi-infinite magnetized plasma slab with a smooth plasma-vacuum interface. Gray curves: propagating bulk and ramp-localized continuum modes. Undamped ITSPWs: red dot-dash line. Damped ITSPW: real $\omega$ (pink dot-dash) and imaginary (green dot-dash; $\gamma/k_p$). Blue stars correspond to the damped mode at $\omega = \omega_{\rm dr} =  0.16 \omega_{p0}, ky = k_p $. The ramp length is $k_p \delta = 0.1$. Rest parameters are same as used in Fig. \ref{Fig4}. (b) A time snapshotof $z-$directional electric field energy at a fixed $z-$plane for damped (blue star in (a)).}
\end{figure}

In Fig.~\ref{Fig5}(a), the dispersion relation for the Ion-Cyclotron Topologically Protected Surface Wave (ITSPW) (represented by the pink dot-dash line) is plotted, illustrating its continuation from the undamped ITSPW (red dot-dash line) into the region of the lower-hybrid continuum. The corresponding damping rate is indicated by the green dot-dash line (top scale). The mode denoted by the star is excited in a 3D particle-in-cell (PIC) simulation, as shown in Fig.~\ref{Fig5}(b). The simulation results clearly demonstrate that the mode undergoes damping as it propagates along the plasma-vacuum interface. Additionally, the green dashed line represents the numerical damping rate, which is highlighted by a blue star in Fig.~\ref{Fig5}(a). Notably, the numerically obtained damping rate aligns well with the theoretical prediction derived from the aforementioned model.

In conclusion, this study demonstrates that the inclusion of finite ion mass gives rise to a second nontrivial low-frequency band gap. The Chern numbers have been computed for all bands, and for the low-frequency bands, they are found to be independent of both the parallel wavenumber $k_z$ and the ion mass. An edge state emerges within the low-frequency band gap for certain nonzero values of $k_z$. The dispersion relation is obtained via a numerical solution of the governing equations for a magnetized two-fluid cold plasma slab.

The undamped ITSPW exists below the ion cyclotron frequency, where no naturally occurring resonance continua are present. Consequently, collisionless damping is not expected in this regime. However, a quasi-edge state has been identified between the ion cyclotron frequency and the second bulk band ($\omega_2$), as previously discussed in Ref.~\cite{Rajawat_arxiv_2022a}. A semi-analytical model has been developed under the limit $\delta l \ll 1/k_p$, where a continuum of lower-hybrid frequencies exists. The edge state undergoes collisionless damping via resonant interaction with the lower-hybrid resonance, thereby transforming into a quasi-edge state.

Furthermore, it is worth noting that since quasi-edge states are confined between $\Omega_i$ and $\omega_2$, they can exist at arbitrary values of $k_z$.   

On the separate note, we do not observe any Alfven resonance bands from our solver. The Alfven resonance may not exist in two-fluid description of a cold plasma and a detailed explanation can be found in \cite{Bellan_pop_1994} and its subsequent comments. 

\bibliography{low_freq_topo.bib}

\begin{acknowledgments}
This work was supported by Air Force of Scientific Research (AFOSR) through Stanford University under MURI Award no. FA9550-21-1-0244. The authors thank the Texas Advanced Computing Center (TACC) for providing HPC resources.
\end{acknowledgments}

\end{document}